\documentclass[aps,prl,twocolumn,showpacs,superscriptaddress,groupedaddress]{revtex4}
\usepackage{graphicx}
\usepackage{color}
\usepackage{chemmacros}
\usepackage{filecontents}
\usepackage[titletoc,toc,title]{appendix}
\begin{document}
\bibliographystyle{nature1}


\title{On the origin of super-diffusive behavior in a class of non-equilibrium systems}

\author{Himadri S. Samanta}\affiliation{Department of Chemistry, University of Texas at Austin, TX 78712}
\author{D. Thirumalai }\affiliation{Department of Chemistry, University of Texas at Austin, TX 78712}
\date{\today}
\begin{abstract}
Experiments and simulations have established that dynamics in a class of living and abiotic systems that are far from equilibrium exhibit super diffusive behavior at long times, which in some cases (for example evolving tumor) is preceded by slow glass-like dynamics. By using the evolution of a collection of tumor cells, driven by mechanical forces and subject to cell birth and apoptosis, as a case study we show theoretically that on short time scales the mean square displacement is sub-diffusive due to jamming, whereas at long times it is super diffusive. The results obtained using stochastic quantization method, which is needed because of the absence of fluctuation-dissipation theorem (FDT), show that the super-diffusive behavior is universal and impervious to the nature of cell-cell interactions. Surprisingly, the theory also quantitatively accounts for the non-trivial dynamics observed in simulations of a model soap foam characterized by creation and destruction of spherical bubbles, which suggests that the two non-equilibrium systems belong to the same universality class. The theoretical prediction for the super diffusion exponent is in excellent agreement with simulations for collective motion of tumor cells and dynamics associated with soap bubbles. 
  \end{abstract}

\maketitle

Collective movement of cells is a pervasive phenomenon in many processes in biology ranging from tissue remodeling that underlies embryonic morphogenesis to wound repair and cancer
invasion~\cite{Christiansen06CR,Friedl95CR,Lecaudey06COCB,Vaughan66JCS,Weijer09JCS,Hanahan11Cell}. 
Consequently, there is considerable interest in understanding the dynamics associated with such processes.
During migration, cells move as sheets, strands, clusters or ducts rather than
individually, and use similar actin- and myosin-mediated protrusions and guidance by
extrinsic chemotactic and mechanical cues just as in the motility of single cells~\cite{Friedl09NRCMB,Friedl04IJDB,Lecaudey06COCB,Roth07CB, Gelimson15PRL}. 
Collective invasion during cancer progression, accompanied by the destruction of tissues and remodeling of the extra cellular matrix, is also important in metastasis \cite{Friedl04IJDB,Friedl09NRCMB,Ilina09JCS,Friedl12NCB}.~The dynamics of these processes are complicated because of an interplay of inter cell adhesive interactions and the biology governing cell birth and apoptosis. The dynamical events involving cell birth and apoptosis implicitly generate  active forces~\cite{Angelini11PNAS,Schotz13JRSI,Ranft10PNAS}, thus driving the systems far from equilibrium. 
How the interplay of death-birth processes and cell-cell interactions in a growing tumor spheroid, poise the cells for effective invasion into the surrounding matrix, is poorly understood.

Complex dynamics in the systems mentioned above manifests itself as caging of a cell by surrounding cells and dynamic heterogeneity-features that are reminiscent of supercooled liquids \cite{Kirkpatrick15RMP}. There are also some surprising departures from glass-like behavior, which is revealed by the super-diffusive behavior on long time scales. For example, experiments on tumor cells invading a collagen matrix~\citep{Valencia15} have shown that at long times (times exceeding the cell division time) the mean square displacement of tumor cells, $<\Delta r^2(t)> \sim t^\alpha$, exhibits super diffusive behavior with $\alpha \approx 1.4 \pm 0.04$. Interestingly, rheology in completely unrelated synthetic materials (foams and mayonnaise) modeled as compressible spherical bubbles, which can be created or destroyed, also exhibit similar behavior. Simulations of such soft glassy materials~\cite{Hwang16NM} show that $<\Delta r^2(t)> \sim t^\alpha$ at long times with $\alpha \approx 1.37 \pm 0.03$.~Both tumor growth and ripening of bubbles are intrinsically non-equilibrium systems because cells (or bubbles) are born as a result of mitosis and also undergo apoptosis. Is there a common mechanism for the origin of super-diffusive behavior in these seemingly unrelated non-equilibrium systems and if so can the long-time universal behavior be explained theoretically? 
\begin{figure}[b]
\vspace{-1.73 in}
	\includegraphics[width=.45\textwidth]{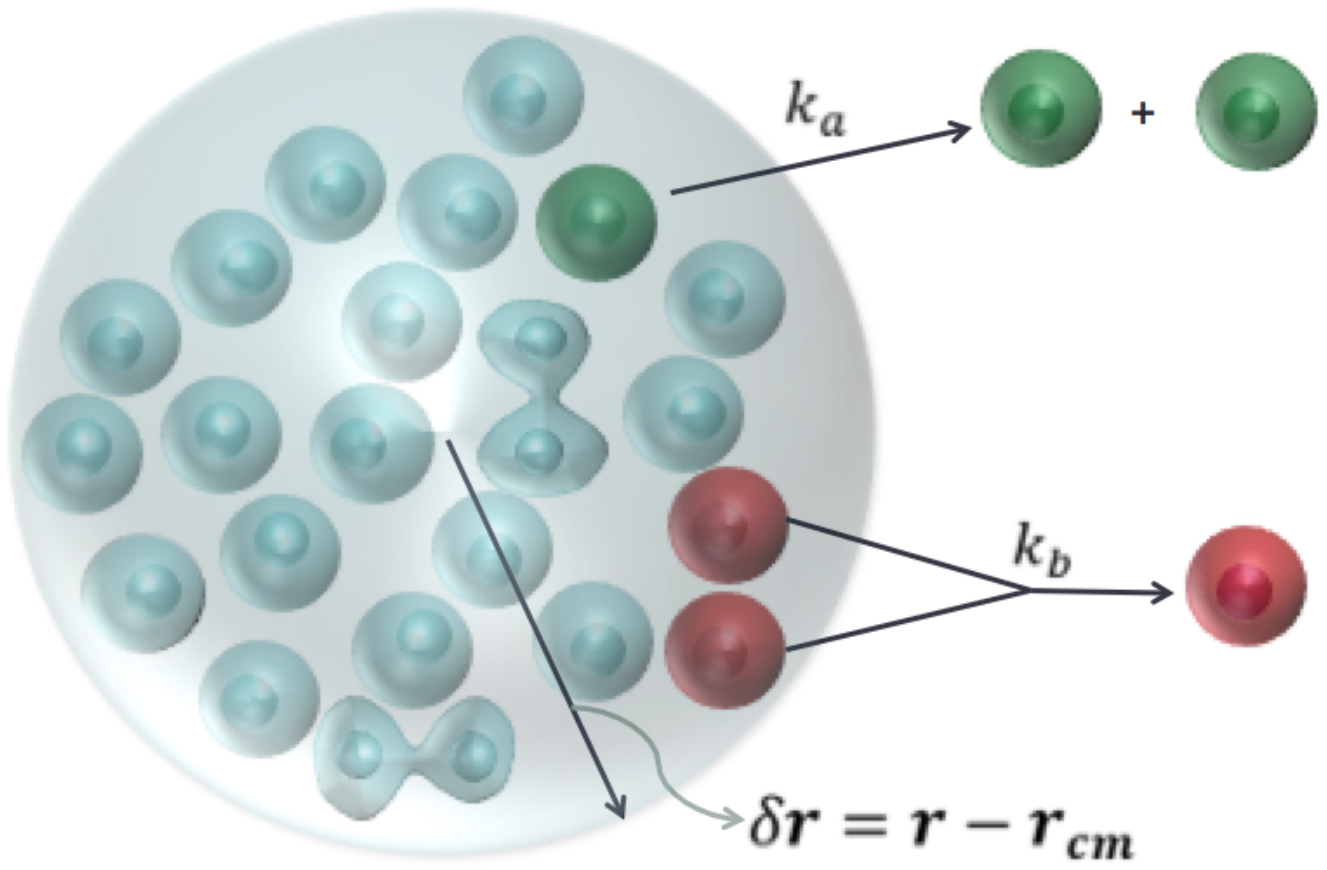}
		\vspace{-1.7 in}
	\caption{Schematic of the dynamics associated with cell birth and death. Cell in green color divides into two cells with rate $k_a$ and cells in red color involve in the cell death process with rate $k_b$. The invasion distance, $\delta {\bf r}= {\bf r}- {\bf r}_{cm}$, is a measure of the extent of penetration of the tumor into the surrounding matrix. }
	\label{fig:rg5}
\end{figure}

Here, we answer the questions posed above in the affirmative by developing a theory to describe the non-equilibrium dynamics of collective cell migration. Cells are modeled as deformable objects interacting with potentials that account for repulsive elastic forces and intercell adhesive attractions due to interactions between cadherins expressed on the cell surface. In addition, the cells could divide at a rate $k_a$, giving rise to daughter cells, and undergo apoptosis, at a rate $k_b$.~Due to the  death-birth processes (Fig. 1), cell number conservation is violated, thus making it difficult to use standard methods to solve the stochastic equations describing the evolution of cell density. A similar scenario arises in the description of dynamics of chemotactic cells, in which cell division and death play an important role \cite{Gelimson15PRL}. Because Gelimson and Golestanian were primarily interested in the long time collective dynamics, they resorted to dynamical renormalization group techniques to investigate the interplay of chemical signaling and cell growth. We follow a different route to study the relevant continuum description of collective behavior of a colony of cells in both the finite as well as in the long time limit, using stochastic quantization method, which was introduced by Parisi and Wu~\cite{Parisi81ES}, in the context of quantum field theory. 

The major results of this study are: (i) The interplay between non-linear terms that determine the intercellular interactions (adhesion and excluded volume repulsion, collectively referred to as mechanical interactions from now on) and death-birth processes are manifested in the dynamics that changes dramatically as the system evolves.~At finite times, mechanical interactions with strong attraction between cells, dominate over the effects cell birth and death,~leading to glassy dynamics. The jammed cells exhibit sub-diffusive motion at the intermediate time scales, where the mean-square displacement, $\langle \Delta r(t) \rangle^{2}$ increases sub-linearly, as $t^{\alpha}$ with $\alpha=0.8$.~(ii) In the long time limit, the consequences of the birth-death processes dominate over the mechanical interactions, resulting in the fluidization of cells. Asymptotically, the cells exhibit super-diffusive motion, with $\langle \Delta r(t) \rangle^{2} \approx t^{\alpha}$ with the value of the universal exponent $\alpha=1.33$, in three dimensions. The theoretical prediction is in excellent agreement with the simulation results~\cite{Abdul18PRX} and a recent {\it in vitro} experiments of the three-dimensional growth of multicellular tumor spheroids~\cite{Valencia15}. (iii) Although the theory is set in the context of tumor growth, the present work also quantitatively describes the complex motion of bubbles in a foam in which bubble formation (birth) and collapse (death) occur.   

We consider the dynamics of a colony of cells in a dissipative environment where inertial effects are negligible.
Each cell experiences  systematic forces arising from mechanical interactions, and a Gaussian random force with white noise spectrum.
The equation of motion for a single cell $i$ is
$
\label{eqmo}
\frac{\partial {\bf r}_i}{\partial t}=-\sum_{j=1}^{N}{\bf \nabla} U({\bf r}_i(t)-{\bf r}_j(t))+\eta_i(t),
$
~where $U$ contains both repulsive interactions with range $\lambda$, and favorable attractive interactions between cells with range $\sigma$, with strengths $v$ and $\kappa$ respectively. We use Gaussian potentials (see the Supplementary Information (SI) for details) in order to obtain analytical solutions. Needless to say that the conclusions would be valid for any short-ranged  $U$.
The Gaussian white noise, satisfies $<\eta_i(t)\eta_j (t')>=2 D\delta_{ij}\delta(t-t')$. Let us consider the evolution of the density function for a single cell $\phi_i({\bf r},t)=\delta[\bf r-{\bf r}_i(t)]$. A closed form of the Langevin equation for the density, $\phi({\bf r},t)=\sum_i \delta[\bf r-{\bf r}_i(t)]$ may be obtained using the approach developed by Dean~\cite{Dean96JPA}. The time evolution of $\phi({\bf r},t)$ is given by,
$
\frac{\partial \phi({\bf r},t)}{\partial t}={\bf \nabla} \cdot \left(\eta({\bf r},t) \phi^{1/2}({\bf r},t)\right)+ {\bf \nabla }\cdot \left(\phi({\bf r},t)\int d{\bf r'} \phi({\bf r'},t){\bf \nabla}U({\bf r-\bf{r'}})\right)+D \nabla^2 \phi({\bf r},t).
$
~We extend the model phenomenologically by adding the source term that describes both cell birth and death as well as a noise term that breaks the cell number conservation.
The line of argument follows from the Doi-Peliti formalism~\cite{Doi76JPAMG,Peliti85,Himadri17PRE}, introduced in the study of reaction-diffusion processes.

The Langevin equation,~for the time-dependent changes in the density, $\phi({\bf r},t)$ is  
\begin{eqnarray}
\label{phi10}
&&\frac{\partial \phi({\bf r},t)}{\partial t}= {\bf \nabla }\cdot \left(\phi({\bf r},t)\int d{\bf r'} \phi({\bf r'},t){\bf \nabla}U({\bf r-\bf{r'}})\right)\\ \nonumber&&+D \nabla^2 \phi({\bf r},t)+k_b \phi(\frac{k_a}{k_b}-\phi)+{\bf \nabla} \cdot \left(\eta({\bf r},t) \phi^{1/2}({\bf r},t)\right)\\ \nonumber
&&+\sqrt{k_a \phi+k_b \phi^2} f_\phi \, ,
\end{eqnarray}
with $f_\phi$ satisfying $<f_\phi({\bf r},t)f_\phi({\bf r'},t')>=\delta({\bf r}-{\bf r}')\delta(t-t')$.
The source term, $g \phi(\phi_0-\phi)$ (third term on the right hand side of Eq.~(\ref{phi10})), arises due to the cell death-birth processes (Fig. 1), with an effective growth rate $g=k_b$, and carrying capacity $\phi_0=\frac{k_a}{k_b}$~\cite{Doering03PA,Gelimson15PRL}. The coefficient~$\sqrt{k_a \phi+k_b \phi^2} $ is the strength of the noise due to number fluctuations, and is a function of density ($\phi$). 

The absence of a fluctuation-dissipation theorem (FDT), due to the generation of active forces, makes this a far from equilibrium problem. Although dynamic renormalization group methods could be used to solve Eq.~(\ref{phi10}) in the  hydrodynamic limit \cite{Gelimson15PRL}, it would not capture the dynamics in the intermediate time regime. Our focus is to study the collective dynamics in a colony of tumor cells in both the intermediate and long time limits. Therefore, we solve Eq.~(\ref{phi10}) by treating the non-linear terms as a perturbation, 
by adopting the stochastic quantization scheme~\cite{Parisi81ES,Himadri06PLA,Himadri06PRE}, which allows us to calculate the form of the MSD in the intermediate as well as the long time limit.

We assume that the density fluctuates around a constant value, which simplifies the multiplicative noise term (last term in Eq.~(\ref{phi10})). We write the density as $\phi({\bf r},t)=\phi_0+\phi_1({\bf r},t)$, 
and perform a linear stability analysis in the Fourier space for the equation describing density fluctuations. 
From the linear stability analysis, we find that the uniform density phase is stable if $(\phi_0 k^2 U({\bf k}) -(k_a-2k_b \phi_0))>0$ (Eq.(S18) in the SI). In this regime, mechanical interactions dominate over the cell birth-death and is the primary determinant of the dynamics of cells.
In the opposite limit, when the active forces due to cell birth-death dominate, the cell colony grows rapidly. There is  
an instability at $(\phi_0 k^2 U({\bf k}) -(k_a-2k_b \phi_0))=0$, signaling a transition from sub-diffusive to super-diffusive motion in the cell dynamics (see below).
 
To anticipate the  consequences of non-linearity, we introduce a change of scale ${\bf r}\rightarrow s {\bf r}$, $\phi \rightarrow s^{\chi} \phi $ and $t\rightarrow s^z t$ where $\chi $ is the exponent corresponding to the cell density fluctuations, and $z$ is the dynamical exponent. The nonlinear term $ (-{\bf q}\cdot {\bf k})U({\bf q})\phi_1({\bf q})\phi_1({\bf k}-{\bf q})$ representing the cell-cell mechanical interactions scales as $s^{2\chi-2}$. The term $b\phi_1({\bf q})\phi_1({\bf k}-{\bf q})$, due to stochastic cell birth-death processes, scales as $s^{2\chi}$. In the hydrodynamic limit ($k\rightarrow 0$ and $t\rightarrow \infty$), nonlinearity due to cell birth-death dominates over mechanical interaction. Therefore, in the hydrodynamic limit, scaling behavior is determined by the death-birth process, which implies that one expects universality in the scaling of the MSD in the long time limit. These conclusions are supported by recent simulation results~\cite{Abdul18PRX}. However, in the intermediate time regime  all the terms contribute to the time dependence of the MSD, $\Delta (t)$. By choosing the strength of the interactions, in such a way that the mechanical interactions dominate over death-birth term (first term in Eq.~(\ref{phi10})), we can calculate $\Delta (t)$ as a function of $t$. 

We now provide a theory in support of the arguments given above.
As stated earlier a major difficulty in studying the problem of collective behavior of cells far from equilibrium is the breakdown of the FDT. Therefore, independent diagrammatic expansions for the response function $<\tilde{\phi}_1\phi_1>$ and the correlation function $< \phi_1\phi_1>$ are necessary. The equilibrium distribution is unknown, and may not exist. Therefore, the averages can be computed only for the statistical noise. The usual analytic route employed in calculating the scaling exponents, is to introduce a response field $\tilde{\phi}_1$, and compute the response function as $<\tilde{\phi}_1 \phi_1>$ and the correlation function as $< \phi_1\phi_1>$. One can obtain the scaling solutions of the relevant problem by using dynamic renormalization group (RG) scheme, as illustrated recently~\cite{Gelimson15PRL}. {The novelty of our theory is that it successfully captures the growing phase of the tumor, which is not easily accessible in the perturbative calculation using the RG scheme~\cite{Gelimson15PRL}. Here, we develop a general theoretical formalism, in which scaling solutions can be obtained by power counting analysis.}

The probability distribution corresponding to the noise term is given by,
\begin{eqnarray}\label{dist}
P(f_{\phi_1}) &\propto & \text{exp}\left[-\int_{{\bf k},w}\frac{1}{2}f_{\phi_1}({\bf k},w)f_{\phi_1}(-{\bf k},-w)\right ]\\ \nonumber 
&&=\text{exp}[ - \frac{1}{2(k_a \phi_0+k_b \phi_0^2)} \int_{{\bf k},w} \mathcal{S}({\bf k},w)].
\end{eqnarray} 
The action functional $\mathcal{S}({\bf k},w)$ may be written in terms of $\phi_1({\bf k},w)$ instead of $f_{\phi_1}({\bf k},w)$,
with the help of Eq.~(\ref{phi10}).
We now exploit the Parisi-Wu stochastic quantization scheme~\cite{Parisi81ES}, and introduce a fictitious time $`\tau_f$' and 
consider all variables to be functions of $\tau_f$ in addition to {\bf k} and $w$. 
The Langevin equation in the $\tau_f$ variable is (see SI for details),
\begin{equation}\label{langefic}
\frac{\partial \phi_1({\bf k},w,\tau_f)}{\partial \tau_f}=-\frac{\delta \mathcal{S}}{\delta \phi_1(-{\bf k},-w,\tau_f)}+f_{\phi_1}({\bf k},w,\tau_f) \, ,
\end{equation}
where $f_{\phi_1}$ satisfies,  $<f_{\phi_1} f_{\phi_1}>=2 \delta(k+k')\delta(w+w')\delta(\tau_f-\tau_f')$.
This ensures that as $\tau_f\rightarrow \infty$, the distribution function will be given by $S({\bf k},w)$ in Eq.~(\ref{dist}) because FDT holds in the $\tau_f$ variable. 
The correlation functions calculated using Eq.~(\ref{langefic}) lead to the physical correlation functions of the original theory (Eq.~(\ref{phi10})) in the $\tau_f \rightarrow \infty$ limit~\cite{Himadri06PRE}. 

In order to obtain the scaling laws for the MSD, it suffices to work at arbitrary $\tau_f$. It follows from Eq.~(\ref{langefic}) that in the absence of the non-linear terms, the Greens function $G^{(0)}$ is given by,
$[G^{(0)}]^{-1}=-i\omega_{\tau_f}+\frac{1}{2(k_a \phi_0+k_b \phi_0^2)}[  \omega^2 +\{ D k^2+\phi_0 k^2 U({\bf k})-(k_a-2k_b\phi_0)\}^2] \, ,
$
~where $\omega_{\tau_f}$ is the frequency corresponding to the fictitious time $\tau_f$. The effect of non-linear terms can be included perturbatively leading to the Dyson's equation,
\begin{equation}\label{green1}
[G]^{-1}=[G^{(0)}]^{-1}+\Sigma({\bf k},\omega, \omega_{\tau_f}),
\end{equation}
{where the self-energy $\Sigma({\bf k},\omega, \omega_{\tau_f})$ contains the non-linear contributions to the bare Greens function. We obtain the following self-consistent equation for the self-energy from the above Greens function equation~\cite{SI2018},
\begin{equation}
\label{scale}
 \Delta\nu k^2=\frac{1}{2\nu k^2} \Sigma({\bf k},\omega, \omega_{\tau_f}),
\end{equation}
with $\nu=D+\phi_0  U({\bf k})$. The two loop contribution from the first order term (containing two $\phi_1$ fields) in Eq. (\ref{langefic}) will contribute to the scaling laws in the intermediate as well as in the long time limit (see below).}

{{\it Sub-diffusive motion}:}
In the spirit of self consistent mode coupling theory, we replace $\nu$ by $\Delta \nu$ in 
the self energy term $\Sigma({\bf k},\omega, \omega_{\tau_f})$. We use $G$ from 
Eq.~(\ref{green1}), and an expression for $C$ follows from the FDT. 
{According to scale transformation, we know $\omega \sim k^z$, $\omega_\tau \sim k^{2z}$, $G \sim k^{-2z}$, $C \sim k^{-4z}$ and the vertex factor $V \sim k^{z+2}$. The self energy term (Fig.~(S1) in the SI), can be written as 
$\Sigma({\bf k},\omega, \omega_{\tau_f})\sim  \int \frac{d^d {\bf k'}}{(2\pi)^d} \frac{d\omega'}{2\pi} \frac{d\omega'_\tau}{2\pi} V V GC$.}
By carrying out the momentum count of $\Sigma({\bf k},\omega, \omega_{\tau_f})$, and 
keeping in mind that $\Delta \nu k^2\sim k^z$, we find that $\Sigma({\bf k},\omega, \omega_{\tau_f})\sim k^{d-z+4}$. 
Using Eq.~(\ref{scale}), we have $k^{z}\sim k^{d-z+2}$, which leads to $z=1+\frac{d}{2}$.

The single cell mean-square displacement behaves as, 
\begin{equation}
<[r(t)-r(0)]^2>\sim t^{2/z}=t^\alpha.
\end{equation}
In 3D, $\alpha=\frac{4}{5}= 0.8$, implying that a labelled cell undergoes sub-diffusive motion, which is one characteristic feature of glassy systems.~{If cell-cell interaction is modeled as $U_1=U_0/\cosh^2(r/a)$ instead of a Gaussian, we obtain $\alpha=\frac{4}{6}= 0.57$, implying sub-diffusive behavior. Although sub-diffusive behavior is preserved at intermediate times, the scaling exponents depend on the form of interaction potential, which shows that the intermediate behavior of $<\Delta r^2 (t)>$ is non-universal. The sub-diffusive behavior is a consequence of jamming of cells. }

{We also investigated how the jamming regime depends on the cell-cell adhesion strength, $\kappa$. The form of the interaction potential is shown in Eq.~(S16) of SI. We define the time dependent order parameter in terms of the function,
$
<Q(t)>\equiv \int dr_1 dr_2 <\phi(r_1, 0) \phi(r_2,t)>\delta(r_1-r_2),
$
~measuring the number of 'overlapping' cells in two configurations separated by a time interval $t$.
In Fourier space, 
\begin{eqnarray}\label{order}
<Q(t)>&=& \int_{{\bf k},w} <\phi_1({\bf k}, w) \phi_1(-{\bf k} ,-w)>e^{i w t} \\ \nonumber
&=& \int_{\bf k}  \frac{1}{\frac{\Sigma({\bf k}) }{\kappa k^2}} \exp[-t \frac{\Sigma({\bf k})}{\kappa k^2}]= \int_{\bf k} \tilde{S}({\bf k},t)
\end{eqnarray}
where the second line is obtained using mode-coupling approximation. The dynamic structure factor $\tilde{S}({\bf k},t)$ decays exponentially (Eq.~(\ref{order})) from which it follows that the relaxation time depends linearly on the adhesion strength $\kappa$. For small value of $\kappa$, $\tilde{S}({\bf k},t)$ decays rapidly and for large $\kappa$, the relaxation time increases substantially leading to stronger caging effect, which results in the extremely slow relaxation of the dynamic structure factor~\cite{Marusyk12NCR,Kirkpatrick15RMP}.  }

{{\it Long time super-diffusion}:} In the long time limit, the effects of non-linearity due to death-birth dominate over mechanical interactions. 
Following the same procedure outlined above, we obtain the self-consistent mode coupling equation of the form, $\Delta \mu=\frac{1}{2\mu} \Sigma({\bf k},\omega, \omega_{\tau_f})$, in the hydrodynamic limit,
with $\mu=(k_a-2k_b\phi_0)$. We now replace $\mu$ by $\Delta \mu$ in the 
self energy term $\Sigma({\bf k},\omega, \omega_{\tau_f})$ (Fig.~(S1) in the SI), use $G$ in
Eq.~(\ref{green1}), and $C$ is calculated using the FDT. The scale transformation for all the variables is the same as before except that the vertex factor $V \sim k^{z}$. By noting that $\Delta \mu \sim k^z$, we find $\Sigma({\bf k},\omega, \omega_{\tau_f})\sim  \int \frac{d^d {\bf k'}}{(2\pi)^d} \frac{d\omega'}{2\pi} \frac{d\omega'_\tau}{2\pi} V V GC\sim k^{d-z}$. The self-consistent equation $\Delta \mu=\frac{1}{2\mu} \Sigma({\bf k},\omega, \omega_{\tau_f})$ produces the dynamic exponent $z=d/2$. Therefore, asymptotically $\alpha =1.33$,
implying that the MSD exponent is greater than unity, which implies that collective motion leads to super-diffusive behavior. The calculated value of $\alpha$ is in excellent agreement with both the value obtained from simulations \cite{Abdul18PRX} and experimental results \cite{Valencia15}.

{\it Invasion distance:} Recently, the movement of the fibrosarcoma cells at the boundary of a growing spheroid pushing against collagen matrix was measured using imaging techniques~\cite{Valencia15}. The dynamics was quantified using the invasion distance (Fig.~(\ref{fig:rg5})), which is defined as the average distance from the center of mass of the tumor to the cells at the periphery, $\delta r(t) =<r_i-r_{CM}>$, where $r_{CM}=(1/N)\sum_{i} r_i$, with $N$ being the number of cells. It was found that $<\delta r(t)> \sim t^{1/z}=t^{\xi}$ with $\xi=0.8$~\cite{Valencia15}. Using our theory we find that $<\delta r(t)> \sim t^{2/3}$. The calculated and measured values of $\xi$ are in fair agreement. If the dynamics were purely diffusive, as would be the case for a homogeneously distributed sample of individual cells, then $\xi$ would be 0.5. The departure from this value is another indication of super-diffusion in this non-equilibrium system.~The time dependent structure factor $\tilde{S}({\bf k},t)$ in this case decays exponentially as $\exp[-t \frac{\Sigma({\bf k})}{\mu }]$, implying that the relaxation time depends linearly on the birth rate \cite{Abdul18PRX}.

{{\it Dynamics of soft material and growing tumor are similar}:} Interestingly, dynamics of certain soft glassy materials and collective migration of cells have a common feature is that the underlying dynamics of these systems are governed by the birth and death processes.
For the soft foam, Hwang and co-workers~\cite{Hwang16NM} used a model for Ostwald ripening for the bubbles, which can be recast as the reaction $X+X\rightarrow X$, that is identical to the apoptosis process used in tumor evolution. This process produces the non-linearity ($k_b \phi^2$) in both the problems. The present theory shows that this non-linear term determines the scaling behavior in the long time limit. From the theory presented above we conclude that both $\Delta r^2(t)$ must have the same scaling behavior. Our theory predicts the general feature that birth-death driven dynamics should lead the super-diffusive behavior with a universal dynamical exponent in the long time limit. Thus, asymptotically MSD scaling is impervious to the interaction details between the constituent objects in the non-equilibrium systems. Based on the calculation of $<\Delta r^2(t)>$ for cells in tumor we surmise that the mean square displacement for bubbles should increase as $t^\alpha$ at long times with the same exponent, $\alpha \approx 1.33$. Remarkably, this value is in accord with the simulation results reported elsewhere \cite{Hwang16NM}.

In summary, using a new theoretical framework, we have provided insights into the dynamics of a colony of tumor cells driven by an interplay of mechanical interactions and stochastic death-birth processes. The breakdown of number conservation, resulting from stochastic death-birth process, makes the dynamics far from equilibrium, characterized by the absence of FDT. The introduction of a fictitious time in which FDT is valid, allows us to calculate the response functions from which the correlation functions can be obtained using the FDT. This new approach greatly simplifies the calculation of the scaling exponents. Non-linear terms in the density evolution equation, arising from mechanical interactions determine the scaling behavior in the intermediate time. Strong cell-cell adhesion interactions lead to the glass-like caging behavior characterized by sub-diffusive motion in the intermediate time. Stochastic death-birth processes determine the scaling in the long time limit, which is independent of the mechanical interactions, as long as they are short-ranged. In the long time limit, the dynamics shows super-diffusive motion, leading to fluidization of the colony of cells. Our theory shows that the universal long time behavior would arise in any systems in which the cells (or particles) are born and undergo apoptosis. These dynamical processes, surely relevant in many biological processes, produce active forces of sufficient magnitude to fluidize the dynamics of jammed cells at long times. It is this mechanism that apparently is also operative in soft glassy materials~\cite{Hwang16NM}, that produces the unexpected super diffusion in this abiotic system. As a consequence of the fundamental similarity between these completely distinct problems, we assert that asymptotically the cells in an evolving tumor and bubbles in a soap foam have precisely the same underlying dynamics at long times. In other words, these non-equilibrium systems belong to the same universality class. It would be most interesting to explore if the mechanism proposed to explain the origin of super-diffusion is present in other non-equilibrium  systems as well. Finally the theory presented here could help us to understand how cancer spreads by invading adjacent tissue involved in metastasis~\cite{Polyak09NRC}.

\bigskip
{\bf Acknowledgments:} This work was supported by the National Science Foundation through NSF Grants No. PHY 17-08128 and No. CHE 16-32756. Additional support was provided by the Welch Foundation through the Collie-Welch Chair with Grant No. F-0019.

\newpage
\begin{widetext}
{
\section{Supplementary Information: On the origin of super-diffusive behavior in a class of non-equilibrium systems }
\subsection{Non-linear term arising from birth-death process}

We consider a minimal model to study the interplay between stochastic cell growth and annihilation process leading to apoptosis, and use it to derive a Langevin type equation for logistic growth. 
We use the Doi-Peliti formalism\cite{Tauber14,doi76JPA,peliti85JP} in order to derive an expression for the density dependence of the noise strength that describes cell number fluctuations. The birth reaction  $X \xrightarrow[]{k_a} X+X$ occurs with the rate constant $k_a$ for each cell, and the backward reaction (annihilation or apoptosis)  $X+X\xrightarrow[]{k_b} X$ occurs with rate $k_b$ (see Fig. 1 in the main text). The master equation for this process is written as, 
\begin{eqnarray}\label{master}
\frac{\partial P(X_i,t)}{\partial t}=&& k_a [(X_i-1)P(X_i-1,t)-X_iP(X_i,t)] \\ \nonumber
&&+\frac{k_b}{V} [X_i(X_i+1)P(X_i+1,t)-X_i(X_i-1)P(X_i,t)],
\end{eqnarray}
where $P(X_i,t)$ is the probability of finding $X_i$ particles at time $t$, and $k_b/V$ is taken to be the apoptosis rate of distinct pairs of cells within the volume $V$. The central idea of the Doi-Peliti formalism~\cite{Tauber14,doi76JPA,peliti85JP} is the introduction of a single vector $|\psi(t)>$, which is a collection of a series of infinite number of $P(X_i,t)$ : 
\begin{equation}
\label{phi}
|\psi(t)>=\sum_{X_i=0}^{\infty} P(X_i,t)|X_i> \, .
\end{equation} 
Using Eq.~(\ref{phi}), the master equation in Eq.~(\ref{master}) can be written in a compact form,
\begin{equation}\label{schrodinger}
\frac{\partial}{\partial t} |\psi(t)>=-L(c^{\dagger},c)|\psi(t)> \, ,
\end{equation}
where
\begin{equation}\label{HH}
L(c_i^{\dagger},c_i)=k_a({c_i^{\dagger}}^2-c_i^{\dagger})c_i +\frac{k_b}{V}(c_i^{\dagger}-{c_i^{\dagger}}^2)c_i^2.
\end{equation}
The Bosonic creation operator $c_i^{\dagger}$ and annihilation operator $c_i$ obey,
\begin{equation}
[c_i,c_i^{\dagger}]\equiv c_i c_i^{\dagger}-c_i^{\dagger} c_i=1,
\end{equation}
where [.,.] is the commutator, and the actions of the creation and annihilation operators for the ket vectors $|n>$ are defined as, $c_i^{\dagger}|X_i>=|X_i+1>$, $c_i|X_i>=X_i|X_i-1>$.

The Schr$\ddot{\text{o}}$dinger like equation (Eq.(\ref{schrodinger})) for the evolution of the state of the system may be integrate to find,
\begin{equation}
\mid \psi (t)> =e^{-L t }\mid \psi (0)>
\end{equation}
with the initial state $| \psi \rangle=e^{\bar{X}_0 \sum_i (c_i^\dagger-1)}|0\rangle$.
The initial configuration for the master equation is an independent Poisson distribution at each site,
\begin{equation}
P(\{X_i\};0)=\Pi_i P_0(X_i)=\Pi_i e^{-\bar{X}_0}
 {X_0}^{-X_i}/X_i!,
\end{equation}
with mean initial input and output concentrations $\bar{X_0}$.

Our goal is to compute averages and correlation functions with respect
to the configurational probability $P(\{X_i\};t)$, which is
accomplished by using the projection state
$\langle\mathcal{P}|=\textless0| \Pi_i e^{c_i}$, for which
$\langle \mathcal{P}|0\rangle=1$ and $\textless \mathcal{P}|
c_i^\dagger=\langle \mathcal{P}|$,
since $[e^{c_i},c_j^\dagger]=e^{c_i}\delta_{ij}$.  The average value
of an observable $A(\{X_i\})$ is,
\begin{equation}
\langle A(t)\rangle=\sum_{\{X_i\}}A(\{X_i\})P(\{X_i\};t),
\end{equation}
from which the statistical average of an observable can be calculated using,
\begin{eqnarray}
\langle A(t)\rangle&=&\langle\mathcal{P}| A(\{c_i^{\dagger},c_i\})| \psi(t)\rangle \\ \nonumber
&=&\langle \mathcal{P}| A(\{c_i^{\dagger},c_i\})e^{-H((\{c_i^{\dagger}\},\{c_i\})t}| \psi(0)\rangle.
\end{eqnarray}

We follow a well-established route in quantum many particle
theory~\cite{Negele88}, and derive a field theory
representation by constructing a path integral equivalent of the time
dependent Schr\"{o}dinger equation (Eq.(\ref{schrodinger})) based on coherent
states~\cite{Tauber14}. These are defined as right eigenstates of the
annihilation operators, $c_i |\alpha_i \rangle=\alpha_i |\alpha_i
\rangle$ , with
complex eigenvalues $\alpha_i$. The coherent states
satisfy $|\alpha_i \rangle= \exp(\frac{1}{2}|\alpha_i|^2+\alpha_i
\alpha_i^\dagger)|0\rangle$, the overlap integral $\textless\alpha_j
|\alpha_i
\rangle=\exp(-\frac{1}{2}|\alpha_i|^2-\frac{1}{2}|\alpha_j|^2+\alpha_j^*
\alpha_i)$, and the completeness relation $\int \Pi_i d^2 \alpha_i
|\{\alpha_i\}\rangle \textless \{\alpha_i\}|=\pi$. After splitting the
temporal evolution (Eq.(\ref{schrodinger})) into infinitesimal increments,
inserting the completeness relation at each time step, and with
additional manipulations, we obtain an expression for the
configurational average,
\begin{equation}
\langle A(t)\rangle \propto \int \Pi_i d\alpha_i d \alpha_i^* 
A(\{\alpha_i\})e^{-\mathcal{S}[\alpha_i^*,\alpha_i]}.
\end{equation}
The exponential statistical weight is determined by the action,
\begin{equation}\label{action}
\mathcal{S}[\alpha_i^*,\alpha_i]=\sum_i \left [
    \int_0^{t_f} \left\{ \alpha_i^* (t) \frac{\partial
      \alpha_i(t)}{\partial t} \right \}
    +L(\alpha_i^*,\alpha) \right ] dt.
\end{equation}
Finally, by taking the continuum limit using $\sum_i \rightarrow
a_0^{-d} \int d^d x$, $a_0$ is a {lattice} constant, $\alpha_i
(t)\rightarrow \phi (x,t)$ and
$\alpha_i (t)\rightarrow a_0^d \phi (x,t)$, the expectation value is represented by a
functional integral,
\begin{equation}
\langle A(t)\rangle \propto \int \Pi_i \mathcal{D}[\phi^*,\phi] A(\{\phi\})e^{-\mathcal{S}[\phi^*,\phi]},
\end{equation}
with an effective action
\begin{equation}\label{Action}
\mathcal{S}[\phi^*,\phi]=\int_0^{t_f}\left[\left\{ \phi^* (t) \frac{\partial \phi(t)}{\partial t} \right \} +L(\phi^*,\phi) \right ] dt.
\end{equation}
 In the  Hamiltonian (Eq.(\ref{HH})), $c^\dagger$ is
 replaced   by   the   field    variable   $\phi^*$, and the $c$ operator  becomes $\phi$.

The action in Eq.(\ref{Action}) encodes the stochastic master equation
kinetics through four independent fields
($\phi^*,\phi$). With this formulation, an immediate
connection can be made to the response functional formulation using
the Janssen-De Dominicis formalism for the Langevin equations
\cite{Dominicis76JPC,Janssen76ZPB}. In this approach, the response
field enters at most quadratically in the pseudo-Hamiltonian, which
may be interpreted as an average over Gaussian white noise. With this
in mind, we apply {the} non-linear Cole-Hopf transformation \cite{Cole51QAM,
  Hopf50CPAM}, in order to obtain the quadratic terms in auxiliary fields,
$\phi^*= e^{\bar{\phi}_I}, \  \phi=e^{-\bar{\phi}_I} \phi_I$,
to the action in Eq.(\ref{Action}). The Jacobian for this variable
transformation is unity, and the local particle density is $\phi^*
\phi =\phi_I$. We obtain the following
Hamiltonian,
\begin{equation}
L=-k_b\bar{\phi} \phi(\frac{k_a}{k_b}-\phi)]
+\bar{\phi}^2[\frac{k_a}{2} \phi+k_b \frac{\phi^2}{2}].
\end{equation}
In the above equation, {the} exponential term has been expanded to
second order. The rate equations is obtained through $\delta
\mathcal{S}/ \delta \bar{\phi} \mid_{\bar{\phi}=0}=0$.  The terms
quadratic in the auxiliary field~$\bar{\phi}$
encapsulate the second moment of the Gaussian white noise with zero
mean.

We arrive at an expression for the action for a colony of tumor cells, governed by the dynamics illustrated in Fig. 1 in the main text, in the continuum description,
\begin{equation}
S[\bar{\phi},\phi]=\int dt \{ \bar{\phi}[\frac{\partial \phi}{\partial t}-k_b \phi(\frac{k_a}{k_b}-\phi)]
+\bar{\phi}^2[\frac{k_a}{2} \phi+k_b \frac{\phi^2}{2}]\} \, .
\end{equation}
The term $k_b \phi(\frac{k_a}{k_b}-\phi)$ gives the source term for cell birth-death. The coefficient of $\bar{\phi}^2$ gives the expression for noise correlation in the Langevin description, which breaks the cell number conservation and plays a crucial role in the dynamical behavior of the collection of cells.

\subsection{Short-range interaction}
To obtain the dynamics of an evolving collection of cells, we use the following simplified form for cell-cell interaction,
\begin{equation}\label{HamiltVH}
{U}({\bf r}(i)-{\bf r}(j))=\frac{v}{(2\pi \lambda^2)^{3/2}} 
e^{-\frac{({\bf r}(i)-{\bf r}(j))^2}{2\lambda^2}}-
\frac{\kappa}{(2\pi \sigma^2)^{3/2}} e^{-\frac{({\bf r}(i)-{\bf r}(j))^2}{2\sigma^2}},
\end{equation}
where $v$ and $\kappa$ are the strengths of excluded volume and attractive interactions, respectively.

\subsection{Density equation}
To simplify the multiplicative noise term (last term in Eq.~(1) in the main text), we assume that the density fluctuates around a constant value.
Hence, we define the density using $\phi({\bf r},t)=\phi_0+\phi_1({\bf r},t)$, and expand Eq.(1) in the main text in $\frac{\phi_1}{\phi_0}$ up to the lowest order in non-linearity. The equation for the density fluctuation becomes,
\begin{eqnarray}\label{rho}
&&\frac{\partial \phi_1({\bf r},t)}{\partial t}=D \nabla^2 \phi_1({\bf r},t)+(k_a-2k_b \phi_0)\phi_1({\bf r'},t) +{\bf \nabla}\cdot \left(\phi_0 \int d{\bf r'} \phi_1({\bf r'},t) \nabla U({\bf r-r'})\right)\\ \nonumber &&+{\bf \nabla}\cdot \left(\phi_1({\bf r'},t)  \int d{\bf r'} \phi_0 \nabla U({\bf r-r'})\right) 
+{\bf \nabla }\cdot \left(\phi_1({\bf r},t)\int d{\bf r'} \phi_1({\bf r'},t){\bf \nabla}U({\bf r-\bf{r'}})\right)\\ \nonumber &&+{\bf \nabla} \cdot \left(\eta({\bf r},t) \phi_0^{1/2}\right)-k_b \phi_1^2+\sqrt{k_a \phi_0+k_b \phi_0^2} f_\phi .
\end{eqnarray}
In Fourier space, the above equation reads,
\begin{eqnarray}\label{LF}
\label{rho12}
\frac{\partial \phi_1({\bf k},t)}{\partial t}=&&-(D k^2+\phi_0 k^2 U({\bf k}) )\phi_1({\bf k})+(k_a-2k_b \phi_0)\phi_1({\bf k})\\ \nonumber &&+\int d{\bf q} (-{\bf q}\cdot {\bf k})U({\bf q})\phi_1({\bf q})\phi_1({\bf k}-{\bf q})
-k_b \int d{\bf q} \phi_1({\bf q})\phi_1({\bf k}-{\bf q})+\eta'({\bf k},t),
\end{eqnarray}
with $<\eta'({\bf k},t)\eta'(-{\bf k},t')>=(k_a \phi_0+k_b \phi_0^2+2D \phi_0 k^2) \delta(t-t')$.\\
The Greens function $G$ is given by,
\begin{equation}\label{green1}
[G]^{-1}=-i \omega+D k^2+\phi_0 k^2 U({\bf k})-(k_a-2k_b \phi_0)+\Sigma({\bf k},\omega) ,
\end{equation}
where, $\Sigma({\bf k},\omega) \sim  \int \frac{d^d {\bf k'}}{(2\pi)^d} \frac{d\omega'}{2\pi}  V V GC\sim  \int \frac{d {\bf k'}}{(2\pi)^d} k'^{d-5}$, showing infrared divergence at the critical dimension $d_c=4$. 
For $d>d_c$, scaling exponents are determined by linear theory and for $d<d_c$, non-trivial exponents are governed by the non-linear terms in Eq.(\ref{LF}). 

\subsection{ Stochastic quantization technique} We now exploit the Parisi-Wu stochastic quantization scheme~\cite{Parisi81ES}, and introduce a fictitious time $`\tau_f$', and 
consider all the variables to be functions of $\tau_f$. 
A Langevin equation in $`\tau_f$' space is,
\begin{equation}\label{langefic}
\frac{\partial \phi_1({\bf k},w,\tau_f)}{\partial \tau_f}=-\frac{\delta \mathcal{S}}{\delta \phi_1(-{\bf k},-w,\tau_f)}+f_{\phi_1}({\bf k},w,\tau_f) \, ,
\end{equation}
with $<f_{\phi_1} f_{\phi_1}>=2 \delta(k+k')\delta(w+w')\delta(\tau_f-\tau_f')$.
Because FDT is valid in the fictitious time it follows that as $\tau_f\rightarrow \infty$, the distribution function will be given by the action $S({\bf k},w)$. The action $S({\bf k},w)$ can be obtained by writing down the probability distribution $P(\eta') \propto \text{exp}\left[-\int \frac{d^d{\bf k}}{(2\pi)^d}\frac{dw}{2\pi}\frac{1}{2}\eta'({\bf k},w)\eta'(-{\bf k},-w)\right ]$ corresponding to the noise term $\eta'$ in Eq.(\ref{rho12}), and  the action $S({\bf k},w)$ in terms of $\phi_1({\bf k},w)$ using of Eq.(\ref{rho12}). The expression for the action $\mathcal{S}$ obtained using Eq.(\ref{LF}) is,\\
$\mathcal{S}=\int \frac{d^d{\bf k}}{(2\pi)^d}\frac{dw}{2\pi}\frac{1}{2}
\{(-iw+D k^2+\phi_0 k^2 U({\bf k})) \phi_1({\bf k})-(k_a-2k_b \phi_0)\phi_1({\bf k})-\int d{\bf q} (-{\bf q}\cdot {\bf k})U({\bf q})\phi_1({\bf q})\phi_1({\bf k}-{\bf q})
+k_b \int d{\bf q} \phi_1({\bf q})\phi_1({\bf k}-{\bf q})\}
\{(iw+D k^2+\phi_0 k^2 U({-\bf k})) \phi_1({-\bf k})-(k_a-2k_b \phi_0)\phi_1({-\bf k})-\int d{\bf q} ({\bf q}\cdot {\bf k})U({\bf q})\phi_1({\bf q})\phi_1({-\bf k}-{\bf q})
+k_b \int d{\bf q} \phi_1({\bf q})\phi_1({-\bf k}-{\bf q})\}$.

\subsection{Langevin equation in the $\tau_f$ variable} With the action given above, we obtain the Langevin equation using Eq. (\ref{langefic}) for $\phi({\bf k},\omega, \tau_f)$,
\begin{eqnarray}\label{langevin}
&&\frac{\partial \phi_1({\bf k},\omega, \tau_f)}{\partial \tau_f}=-\frac{1}{(k_a \phi_0+k_b \phi_0^2+D k^2)}[  \omega^2 +\{Dk^2+ \phi_0 k^2 U({\bf k})+k_a\}^2] \phi_1({\bf k},\omega,\tau_f)\\ \nonumber && -\frac{1}{(k_a \phi_0+k_b \phi_0^2+D k^2)} \int_{{\bf k'},\omega'}
\left[\{i \omega+Dk^2+\phi_0 k^2 U({\bf k})+k_a\} \{(-{\bf k'} \cdot {\bf k}) U({\bf k'})-k_b\}\right.\\ \nonumber && \left.+ \{i \omega'+Dk'^2+\phi_0 k'^2 U({\bf k'})+k_a\} \{(-{\bf k'}\cdot {\bf k}) U(-{\bf k})-k_b\} \right. \\ \nonumber  &&\left.+\{i \omega'+Dk'^2+\phi_0 k'^2 U({\bf k'})+k_a\}  \{(-{\bf k'} \cdot ({\bf k}-{\bf k'})) U({\bf k}-{\bf k'})-k_b\}\right]\\ \nonumber 
&&\phi_1({\bf k'},\omega') \phi_1({\bf k}-{\bf k'},\omega-\omega')+f_\phi({\bf k},\omega,\tau_f)+ \text{higher order terms} \, .
\end{eqnarray}
\subsection{Greens function for density equation}
The Greens function $G$ from the above equation is given by,
\begin{equation}\label{green1}
[G]^{-1}=[G^{(0)}]^{-1}+\Sigma({\bf k},\omega, \omega_{\tau_f}) 
\end{equation}
\begin{figure}[b]
	\includegraphics[width=.450\textwidth]{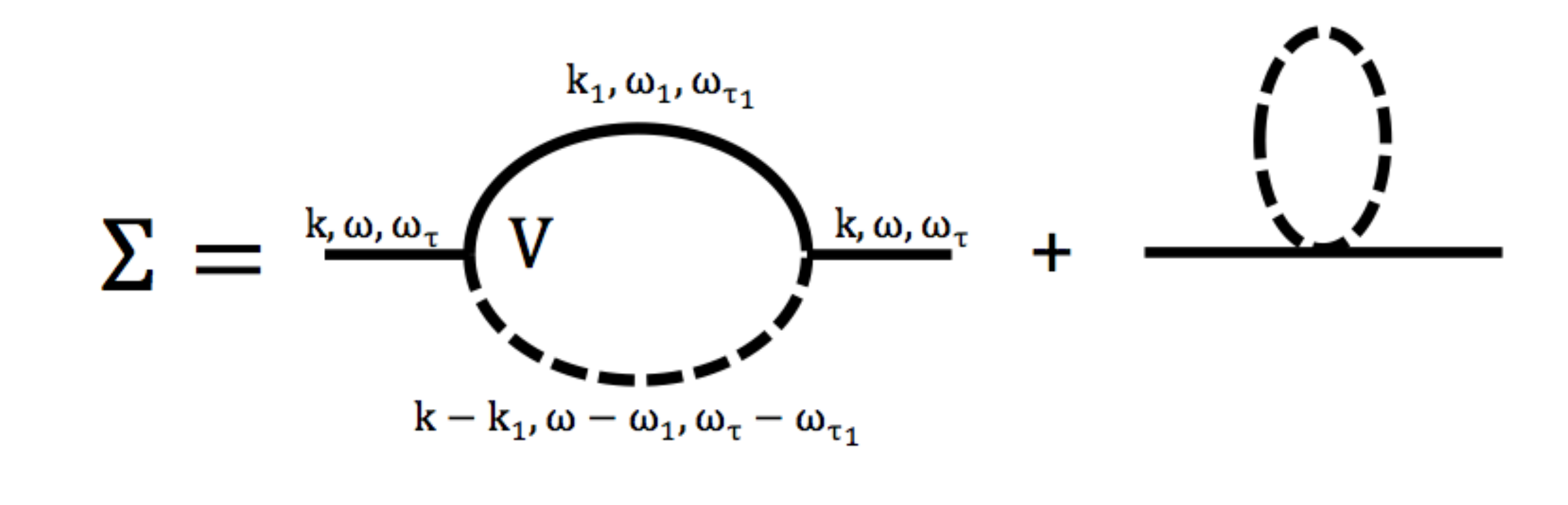}
	\caption{Dashed line indicates the correlation function ($G_0 G_0^*$) and solid line indicates the response function ($G_0$).  The self-energy term ($\Sigma$) is obtained by contracting the two $\phi_1$ fields. First term is the two loop contribution from the first order term (contains two $\phi_1$ fields) in the fictitious time equation in Eq.(\ref{langevin}). The second one gives the one loop contribution from second order term (contains three $\phi_1$ fields). }  
	\label{fig:rg5}
\end{figure}
We are mainly interested in the behavior of $\Sigma({\bf k},\omega, \omega_{\tau_f})$ when expanded to second order in non-linearity. The contributions arise from two sources (1) 
a one-loop contribution from the second order term (containing three $\phi_1$ fields) in Eq.~(\ref{green1}) (second term in Fig.~(S1)) and (2) a two-loop contribution from the first order term ( containing two $\phi_1$ fields)  in Eq. (\ref{green1}) (first term in Fig.~(S1)).  The  contribution arising from the term containing three $\phi_1$ fields, in Eq. (\ref{langevin}) can 
be readily obtained by contracting two of the $\phi_1 $ fields. The  second order contribution due to the one loop contribution in 
Eq. (\ref{green1}) does not have any new momentum dependence.  Hence, it is the second-order contribution (second term in Fig.(\ref{fig:rg5})) arising from the two-loop contribution in Eq. (\ref{green1}) which is  relevant.
The correlation function is given by the FDT as $C=\frac{1}{\omega_{\tau_f}} \text{Im}G$.  
With these observations, Eq. (\ref{green1}) can be written as,
\begin{equation}\label{green2}
[G]^{-1}({\bf k},\omega,\omega_{\tau_f})=-i\omega_{\tau_f}+\frac{1}{2(D_0)}[  \omega^2 ]+\frac{1}{2(\bar{D})}[\nu_{eff}^2 k^4] \, ,
\end{equation}
where $D_0=k_a \phi_0+k_b \phi_0^2$ and $\bar{D}$ is defined as,
\begin{equation}\label{green3}
\frac{1}{2(\bar{D})}[ \nu_{eff}^2 k^4]=\frac{1}{2(D_0)}(\nu k^2 )^2+\Sigma({\bf k},\omega, \omega_{\tau_f})
\end{equation}
with $\nu=D+\phi_0  U({\bf k})$. In the intermediate time, the strength of the interactions is such that $\phi_0 k^2 U({\bf k})$ dominates over $(k_a-2k_b\phi_0)$. We obtain Eq.(\ref{green3}) by neglecting the term $(k_a-2k_b\phi_0)$ in the Green's function equation (Eq.(\ref{green2})) in the finite time regime.
Expanding $\nu_{eff}$, $\bar{D}$ about $\nu$ and $D_0$, respectively, and noting that the renormalization of $\nu $ dominates, we write,
\begin{eqnarray}
\label{scale}
&&\nu_{eff} k^2 \simeq \nu k^2 +\frac{1}{2\nu k^2} \Sigma({\bf k},\omega, \omega_{\tau_f}),\\ \nonumber && \text{or}~ ~\Delta\nu k^2=\frac{1}{2\nu k^2} \Sigma({\bf k},\omega, \omega_{\tau_f}).
\end{eqnarray}
\subsection{The expression for $\Sigma({\bf l},\omega, \omega_{\tau_f})$}
\begin{eqnarray}\label{selfenergy}
&&\Sigma({\bf k},\omega, \omega_{\tau_f})=\frac{2}{(k_a \phi_0+k_b \phi_0^2+Dk^2)^2} \int \frac{d^d {\bf k'}}{(2\pi)^d} \frac{d\omega'}{2\pi} \frac{d\omega'_\tau}{2\pi}   \left[\{i \omega+Dk^2+\phi_0 k^2 U({\bf k})+k_a\}\right.\\ \nonumber && \left. \{(-{\bf k'} \cdot {\bf k}) U({\bf k'})-k_b\}+ \{i \omega'+Dk'^2+\phi_0 k'^2 U({\bf k'})+k_a\} \{(-{\bf k'}\cdot {\bf k}) U(-{\bf k})-k_b\} \right. \\ \nonumber  &&\left.+\{i \omega'+Dk'^2+\phi_0 k'^2 U({\bf k'})+k_a\}  \{(-{\bf k'} \cdot ({\bf k}-{\bf k'})) U({\bf k}-{\bf k'})-k_b\}\right] \\ \nonumber
&&\left[\{i \omega+Dk^2+\phi_0 k^2 U({\bf k})+k_a\}\right. \left. \{(-{\bf (k-k')} \cdot {\bf k}) U({\bf k-k'})-k_b\}+\right. \\ \nonumber  &&\left. \{i (\omega-\omega')+D({\bf k-k'})^2+\phi_0 ({\bf k-k'})^2 U({\bf k-k'})+k_a\} \{(-({\bf k-k'})\cdot {\bf k}) U(-{\bf k})-k_b\} \right. \\ \nonumber  &&\left.+\{i  (\omega-\omega')+D({\bf k-k'})^2+\phi_0 ({\bf k-k'})^2 U({\bf k-k'})+k_a\}  \{(-({\bf k-k'}) \cdot ({\bf k'})) U({\bf k'})-k_b\}\right]\\ \nonumber  &&G({\bf k'},\omega', \omega'_\tau) C({\bf k-k'},\omega-\omega',\omega_{\tau_f}-\omega'_\tau)
\end{eqnarray}

\subsection{Effective diffusion coefficient }
The emergence of super diffusion may be rationalized by considering movement of a labeled cell as a diffusive process with an effective time dependent diffusion coefficient.
In the spirit of mode-coupling theory, we write $D_{eff} k^2 \sim k^z$, where $D_{eff}$ is the effective diffusion coefficient of the cell. In the real time, $D_{eff}$ scales as $t^{\frac{2-z}{z}}$. 
Using Langevin equation of the form $\dot{y}=\sqrt{D_{eff}(t)} \eta_y$, where $<\eta_y(t)\eta_y(t')>=2\delta(t-t')$, we obtain the mean square displacement, $<\Delta y^2>\sim \int D_{eff}(t) dt \sim t^{2/z}$. 

In homogeneous state, the evolution of cells is given by,
\begin{equation}\label{homogeneous}
\frac{\partial \phi_1({\bf r},t)}{\partial t}=D \nabla^2 \phi_1({\bf r},t)+{\bf \nabla} \cdot \left(\eta({\bf r},t) \phi_0^{1/2}({\bf r},t)\right).
\end{equation}

We assume that Eq.(\ref{homogeneous}) is invariant under the scale transformations, ${\bf r}\rightarrow s {\bf r}$, $\phi \rightarrow s^{\chi} \phi $ and $t\rightarrow s^z t$ where $\chi $ is the exponent corresponding the cell density fluctuations, and $z$ is the dynamical exponent. With these transformation, Eq.(\ref{homogeneous}) becomes
\begin{equation}\label{homogeneous1}
\frac{\partial \phi_1({\bf r},t)}{\partial t}=D s^{z-2} \nabla^2 \phi_1({\bf r},t)+s^{-d/2+z/2-\chi-1} {\bf \nabla} \cdot \left(\eta({\bf r},t) \phi_0^{1/2}({\bf r},t)\right).
\end{equation}

To find the critical exponents $z$ and $\chi$, we require that Eq.(\ref{homogeneous}) must be invariant under the scale transformations. Thus, to ensure scale invariance, each term on the rhs of Eq.(\ref{homogeneous1}) must be independent of $s$, which implies that $z=2$ and $\chi=-d/2$. 
Under these conditions, the cells undergo normal diffusion with MSD$\sim t$ . 

In the growing phase, $\phi_1({\bf r},t)$ satisfies,
\begin{eqnarray}\label{homogeneous2}
\frac{\partial \phi_1({\bf r},t)}{\partial t}&=&D \nabla^2 \phi_1({\bf r},t)+(k_a-2k_b \phi_0)\phi_1({\bf r'},t)-k_b \phi_1^2  \\ \nonumber &&+{\bf \nabla} \cdot \left(\eta({\bf r},t) \phi_0^{1/2}({\bf r},t)\right)+\sqrt{k_a \phi_0+k_b \phi_0^2} f_\phi,
\end{eqnarray}
which is obtained by neglecting interaction between cells. 

Using the same scale transformation as before, we obtain,
\begin{eqnarray}\label{homogeneous3}
&&\frac{\partial \phi_1({\bf r},t)}{\partial t}=D s^{z-2} \nabla^2 \phi_1({\bf r},t)+s^{z}(k_a-2k_b \phi_0)\phi_1({\bf r'},t)-s^{\chi+z}k_b \phi_1^2 \\ \nonumber &&+s^{-d/2+z/2-\chi-1} {\bf \nabla} \cdot \left(\eta({\bf r},t) \phi_0^{1/2}({\bf r},t)\right)+s^{-d/2+z/2-\chi}\sqrt{k_a \phi_0+k_b \phi_0^2} f_\phi.
\end{eqnarray}
To ensure scale invariance, one would expect that the rhs of Eq.(\ref{homogeneous3}) must be independent of $s$.
However, this procedure provides five scaling relations for two exponents $z$ and $\chi$, thereby overdetermining them. In order to get the correct values of exponents, the coefficients must also change under scaling. 
Using stochastic quantization scheme mentioned in the main text, we find $z=3/2$ in the long time limit.
The effective diffusion coefficient $D_{eff}$ scales as $t^{1/3}$, thereby, MSD scales as $t^{4/3}$, implying that at long times the motion is  super-diffusive.

}
\end{widetext}

\end{document}